# Student difficulties with determining expectation values in quantum mechanics

Chandralekha Singh and Emily Marshman

*Department of Physics and Astronomy, University of Pittsburgh, Pittsburgh, PA, 15260*

**Abstract.** The expectation value of an observable is an important concept in quantum mechanics. However, upper-level undergraduate and graduate students in physics have both conceptual and procedural difficulties when determining the expectation value of physical observables, especially when using Dirac notation. To investigate these difficulties, we administered free-response and multiple-choice questions and conducted individual interviews with students. Here, we discuss the analysis of data on student difficulties when determining the expectation value.

## I. INTRODUCTION AND BACKGROUND

Learning quantum mechanics (QM) is challenging [1-4]. Investigations of student difficulties in QM are important for developing curricula and pedagogies that help students develop a solid grasp of QM [5-9]. However, few prior research studies have focused on the conceptual and procedural difficulties upper-level undergraduate and graduate students have with expectation values of physical observables when using Dirac notation, a compact and convenient notation used extensively in QM.

Here we discuss an investigation of difficulties that upper-level undergraduate and graduate students have with the expectation values of observables when making use of Dirac notation in courses in which this notation was used extensively. The expectation value is the average value of an observable when measurements of that observable are made on a large number of identically prepared quantum systems. It is used frequently in QM since measurement outcomes are probabilistic rather than deterministic. For each observable $Q$, there is a Hermitian operator $\hat{Q}$. When the quantum system is in a state $|\Psi\rangle$ and an observable $Q$ is measured in an experiment, one obtains an eigenvalue of $\hat{Q}$. Therefore, the expectation value $\langle\Psi|\hat{Q}|\Psi\rangle$ in a given quantum state $|\Psi\rangle$ can be found by summing the probability of measuring a particular eigenvalue of $\hat{Q}$ multiplied by that eigenvalue over all possible measurement outcomes. Ensuring that students conceptually understand the meaning of expectation value and develop proficiency in calculating it is important.

If the states $\{|q_n\rangle, n = 1,2,3 \ldots \infty\}$ form a complete set of eigenstates of a Hermitian operator $\hat{Q}$ corresponding to an observable $Q$ with non-degenerate discrete eigenvalues $q_n$ (i.e., $\hat{Q}|q_n\rangle = q_n|q_n\rangle$), one can find the expectation value of the observable $Q$ in a generic state $|\Psi\rangle$ in terms of the eigenstates and eigenvalues of $\hat{Q}$ by expanding $|\Psi\rangle$ as a linear superposition of the eigenstates of the operator $\hat{Q}$. A generic state $|\Psi\rangle$ can be written as a linear superposition of the eigenstates of an operator $\hat{Q}$ with discrete eigenvalues $q_n$ as $|\Psi\rangle = \sum_n |q_n\rangle\langle q_n|\Psi\rangle = \sum_n c_n|q_n\rangle$, with $c_n = \langle q_n|\Psi\rangle$. Here, $c_n = \langle q_n|\Psi\rangle$ is the projection of the state $|\Psi\rangle$ along an eigenstate $|q_n\rangle$ of the operator $\hat{Q}$ with eigenvalue $q_n$ and $|c_n|^2 = |\langle q_n|\Psi\rangle|^2$ is the probability of measuring $q_n$. Using this, the expectation value can be found as follows:
$\langle\Psi|\hat{Q}|\Psi\rangle = \langle\Psi|\hat{Q}\sum_n c_n|q_n\rangle = \sum_n c_n\langle\Psi|\hat{Q}|q_n\rangle = \sum_n c_n\langle\Psi|q_n|q_n\rangle = \sum_n q_n c_n\langle\Psi|q_n\rangle = \sum_n q_n c_n c_n^* = \sum_n q_n|c_n|^2$.

Another approach for finding the expectation value of observable $Q$ is to insert the identity operator $\hat{I}$ in terms of a complete set of eigenstates of the operator $\hat{Q}$, i.e., $\hat{I} = \sum_n |q_n\rangle\langle q_n|$, into the expression for the expectation value:
$\langle\Psi|\hat{Q}|\Psi\rangle = \langle\Psi|\hat{Q}\hat{I}|\Psi\rangle = \langle\Psi|\hat{Q}\sum_n|q_n\rangle\langle q_n|\Psi\rangle = \sum_n \langle\Psi|q_n|q_n\rangle c_n = \sum_n q_n c_n^* c_n = \sum_n q_n|c_n|^2$.

Moreover, if the states $\{|q\rangle\}$ are a complete set of eigenstates of $\hat{Q}$ with continuous eigenvalues $q$ (i.e., $\hat{Q}|q\rangle = q|q\rangle$) and the identity operator in terms of the eigenstates of $\hat{Q}$ is $\hat{I} = \int_{-\infty}^{+\infty} |q\rangle\langle q|dq$, using a very similar approach to that used for the case in which the eigenvalue spectrum of $\hat{Q}$ is discrete, the expectation value of $Q$ in state $|\Psi\rangle$ in terms of the eigenstates $|q\rangle$ and eigenvalues $q$ is $\langle\Psi|\hat{Q}|\Psi\rangle = \int_{-\infty}^{+\infty} q\,|\langle q|\Psi\rangle|^2 dq$.

## II. METHODOLOGY

Student difficulties were investigated by administering multiple-choice and open-ended questions to upper-level undergraduate (UG) and graduate (G) students in QM courses after traditional instruction in relevant concepts. We observed difficulties on these questions which were administered on in-class quizzes and exams. The UG students were enrolled in a junior/senior level QM course and the G students were enrolled in a first year core graduate QM course. Table I lists the questions that were administered to students as part of this investigation. The multiple-choice question was administered to 184 upper-level UG students after traditional instruction as part of a quiz at four U.S. universities (see Question Q1 in Table I). The open-ended quiz and exam questions were administered to UG and G students after traditional instruction in QM at the University of Pittsburgh over several years (see questions Q2 and Q3 in Table I). The number of students answering the open-ended questions Q2 and Q3 is different in Table I because, in some of the years, UG students were not given question Q3 and we did not include students in data analysis if they left a question blank (e.g., Q2 has 65 G and Q3 has 62 G because three graduate students left Q3 blank). Since the performance on quizzes and exams were comparable, we present consolidated data here. Student performance and difficulties were similar in different years.

**Table I.** Questions involving expectation value that were administered to students and the number of students ($N$) answering the questions. The correct answer is bolded.

| Questions | $N$ |
|---|---|
| **Q1.** Suppose $\{|q_n\rangle, n = 1,2,3 \ldots \infty\}$ forms a complete set of orthonormal eigenstates of an operator $\hat{Q}$ corresponding to a physical observable with non-degenerate eigenvalues $q_n$. $\hat{I}$ is the identity operator. Choose all of the following statements that are correct.<br>(1) $\sum_n |q_n\rangle\langle q_n| = \hat{I}$<br>(2) $\langle \Psi|\hat{Q}|\Psi\rangle = \sum_n q_n |\langle q_n|\Psi\rangle|^2$<br>(3) $\langle \Psi|\hat{Q}|\Psi\rangle = \sum_n q_n \langle q_n|\Psi\rangle$<br>A. 1 only, B. 2 only, C. 3 only, **D. 1 and 2 only**<br>E. 1 and 3 only | 184 UG |
| **Q2.** $|\Psi\rangle$ is a generic state of a quantum system. The states $\{|q_n\rangle, n = 1,2,3 \ldots \infty\}$ are eigenstates of an operator $\hat{Q}$ corresponding to a physical observable with discrete eigenvalues $q_n$. Find the expectation value of $Q$ for state $|\Psi\rangle$ using a basis of eigenstates $|q_n\rangle$ and eigenvalues $q_n$. Show your work. | 127 UG<br>65 G |
| **Q3.** $|\Psi\rangle$ is a generic state of a quantum system. The states $\{|q\rangle\}$ are eigenstates of $\hat{Q}$ with continuous eigenvalues $q$. Find the expectation value of $Q$ for state $|\Psi\rangle$ using a basis of eigenstates $|q\rangle$ and eigenvalues $q$. Show your work. | 32 UG<br>62 G |

The open-ended questions were graded using rubrics which were developed by the two investigators together. A subset of the open-ended questions was graded separately by the investigators. After comparing the grading of the open-ended questions, the investigators discussed any disagreements in grading and resolved them with a final inter-rater reliability of better than 95%.

Student difficulties were also investigated by conducting individual interviews with 23 upper-level UG and G student volunteers enrolled in the QM courses (not necessarily the same students who answered the written questions). The individual interviews employed a think-aloud protocol to better understand the rationale for students' written responses. During the semi-structured interviews, we asked students to "think aloud" while answering the questions. Students first read the questions on their own and answered them without interruptions except that they were prompted to think aloud if they were quiet for a long time. After students had finished answering a question to the best of their ability, we asked them to further clarify issues that they had not clearly addressed earlier while thinking aloud.

Students' reasoning on questions in interviews were used as a guide to generate categories of difficulties and student responses on open-ended questions were coded into categories of difficulties. A subset of student responses on the open-ended questions were coded to determine categories of difficulties by two of the researchers separately. After comparing codes, any disagreements were discussed until full agreement was reached.

### III. FINDINGS

**Difficulty identifying the correct expression for expectation value:** Table II shows that in response to question Q1, only 45% of the students selected the correct answer option D and correctly recognized that the identity operator is $\sum_n |q_n\rangle\langle q_n| = \hat{I}$ and the expectation value of $Q$ is $\langle \Psi|\hat{Q}|\Psi\rangle = \sum_n q_n |\langle q_n|\Psi\rangle|^2$. Table II also shows that 57% of the students selected options B or D that included statement 2. Interviews shed light on why 31% of the students selected the incorrect statement (3), $\langle \Psi|\hat{Q}|\Psi\rangle = \sum_n q_n \langle q_n|\Psi\rangle$, and will be discussed later in this section.

For question Q2, a response was considered correct if the student inserted the identity operator, used an expansion of the generic state $|\Psi\rangle = \sum_n c_n |q_n\rangle$ where $c_n = \langle q_n|\Psi\rangle$, or conceptually reasoned that the expectation value is the sum of all eigenvalues of $\hat{Q}$ multiplied by the probability of obtaining that eigenvalue to obtain the correct final answer. Table II shows that only 29% of the UG students and 52% of the G students were able to obtain a correct expression for the expectation value of $Q$ for state $|\Psi\rangle$ using a basis of eigenstates $|q_n\rangle$ and eigenvalues $q_n$.

Question Q3 was graded using the same rubric as for the discrete case in question Q2. Table II shows that only 22% of the UG students and 53% of the G students were correctly able to obtain an expression for expectation value $\hat{Q}$ for state $|\Psi\rangle$ using a basis of eigenstates $|q\rangle$ and eigenvalues $q$.

Below, we summarize the common conceptual and procedural difficulties involving the expectation value that were observed in written responses and interviews:

**Failing to reason about the expectation value conceptually**: In interviews, students were asked to determine the expectation value and describe conceptually what the expectation value means. Very few students reasoned conceptually that the expectation value is the average of a large number of measurements on identically prepared systems to determine that $\langle \Psi|\hat{Q}|\Psi\rangle = \sum_n q_n |\langle q_n|\Psi\rangle|^2$. Most students used a formal approach to evaluate the expectation value. While some students followed correct procedures such as inserting the identity operator in terms of the eigenstates of the operator or expanding the generic state $|\Psi\rangle$ as a linear superposition of the eigenstates of the operator, many students who tried to use these methods got lost along the way. The fact that so few students were able to reason conceptually about how to determine the expectation value points to the fact that even upper-level UG and G students often prefer "plug and chug" methods as opposed to developing a coherent conceptual understanding that can facilitate the use of the simpler conceptual approaches (which are significantly less prone to error). G students were more facile in using the identity operator to determine the expectation value than UG students. Therefore, the percentages of UG and G students answering Q2 and Q3 correctly in Table II is very different. However, written responses and interviews with UG and G students suggest that many of them did not realize that the expectation value is the average of a large number of measurements on identically prepared systems and they could have reduced their chances of making a procedural mistake if they had used a conceptual approach to find the expectation value.

**Table II.** Percentages of students answering questions related to the expectation value. Percentages of students providing the correct answers are bolded.

| Q1 | A (12%), B (12%), C (5%), **D (45%)**, E (26%) |
|---|---|
| Q2 | **29%** UG students, **52%** G student |
| Q3 | **22%** UG students, **53%** G students |

**Either incorrectly claiming that the operator $\hat{Q}$ acting on $|\Psi\rangle$ yields, e.g., $\hat{Q}|\Psi\rangle = q_n|q_n\rangle$ or $\hat{Q}|\Psi\rangle = q_n|\Psi\rangle$ or arbitrarily replacing $|\Psi\rangle$ with $|q_n\rangle$ (or $|\Psi\rangle$ with $|q\rangle$):** When evaluating $\langle\Psi|\hat{Q}|\Psi\rangle$, some students wrote incorrect expressions for the operator $\hat{Q}$ acting on state $|\Psi\rangle$, e.g., $\hat{Q}|\Psi\rangle = q_n|q_n\rangle$, or $\hat{Q}|\Psi\rangle = q_n|\Psi\rangle$ because they incorrectly reasoned that an operator $\hat{Q}$ acting on a generic state $|\Psi\rangle$ will yield an eigenstate and/or eigenvalue of the operator $\hat{Q}$. This confusion was often due to conceptual difficulty with quantum measurement. In particular, interviewed students with this type of response often incorrectly claimed that an operator acting on a generic state ($\hat{Q}|\Psi\rangle$) describes the measurement process and the right hand side of the equation, e.g., $\hat{Q}|\Psi\rangle = q_n|q_n\rangle$, is the "outcome" of the measurement process [4,8]. For example, one student reasoned: "$\hat{Q}|\Psi\rangle = q_n|q_n\rangle$ because by generalized statistical interpretation, an operator acting on a general state will yield an eigenvalue of that operator with probability $|\langle\Psi|q_n\rangle|^2$." He then wrote: $\langle\Psi|\hat{Q}|\Psi\rangle = \langle\Psi q_n|q_n\rangle = q_n\langle\Psi|q_n\rangle$. This type of difficulty has been observed in other contexts as well [4,8].

Also, some students inappropriately interchanged the states $|\Psi\rangle$ and $|q_n\rangle$ when finding the expectation value. Interviews and written responses suggest that instead of recalling that a generic state $|\Psi\rangle$ can be written as a linear superposition of a complete set of eigenstates of an operator, some students thought that $|\Psi\rangle$ can be written as an eigenstate of $\hat{Q}$ when finding the expectation value of $Q$. For example, some students correctly wrote $\langle\Psi|\hat{Q}|\Psi\rangle$ and then arbitrarily replaced the generic state $|\Psi\rangle$ with the eigenstate $|q_n\rangle$. One student who stated that $|\Psi\rangle = |q_n\rangle$ in this context wrote: $\langle\Psi|\hat{Q}|\Psi\rangle = \sum_{i=1}^{n}\langle q'_n|\hat{Q}|q_n\rangle = \sum_{i=1}^{n} q_n \langle q'_n|\hat{Q}|q_n\rangle = \sum_{i=1}^{n} q_n \delta(q_n' - q_n) = \sum_{i=1}^{n} q_n$. In addition to not understanding how a generic state differs from an eigenstate of $\hat{Q}$ and replacing $|\Psi\rangle$ with $|q_n\rangle$, this student (and many others) made several procedural mistakes. For example, without justification, the student introduced a summation over index $i$ going from 1 to $n$ but the index $i$ is never used in the expression he was summing over (he summed over $i$ but used the index $n$ when writing the eigenstate $|q_n\rangle$). Moreover, instead of using a Kronecker delta, he used a Dirac delta function, which diverges when $q'_n = q_n$. Another student, who arbitrarily replaced the generic state $|\Psi\rangle$ with $|q_n\rangle$, wrote $\langle\hat{Q}\rangle = \langle\Psi|\hat{Q}|\Psi\rangle = \sum_{n=1}^{\infty} q_n \langle q_n|q_n\rangle$. This student also introduced a sum (although it is over the index $n$) but did not justify where it came from. Interviews suggest that at least some students who incorrectly replaced $|\Psi\rangle$ with $|q_n\rangle$ introduced a summation because they remembered that the expectation value of an observable involves a summation.

**Table III.** Percentages of students (out of those who attempted to answer the question) who displayed various difficulties with the expectation value.

| Expectation value of an observable $Q$ (the corresponding operator $\hat{Q}$ has discrete eigenvalues $q_n$) | | |
|---|---|---|
| | UG (N=127) | G (N=65) |
| Writing $\hat{Q}|\Psi\rangle = q_n|q_n\rangle$ or $\hat{Q}|\Psi\rangle = q_n|\Psi\rangle$ or replacing $|\Psi\rangle$ with $|q_n\rangle$ | 19% | 14% |
| Incorrect expansion of $|\Psi\rangle$ | 3% | 5% |
| Incorrect expression for expectation value | 7% | 3% |
| Inserting the identity operator but getting lost along the way | 6% | 5% |
| **Expectation value of an observable $Q$ (the corresponding operator $\hat{Q}$ has continuous eigenvalues $q$)** | | |
| | UG (N=32) | G (N=62) |
| Writing $\hat{Q}|\Psi\rangle = q|q\rangle$ or $\hat{Q}|\Psi\rangle = q|\Psi\rangle$ or replacing $|\Psi\rangle$ with $|q\rangle$ | 38% | 20% |
| Incorrect expression for expectation value | 13% | 5% |
| Attempting to use $\hat{I}$ but getting lost along the way | 13% | 8% |

Students who claimed that the operator $\hat{Q}$ acting on $|\Psi\rangle$ yields, e.g., $\hat{Q}|\Psi\rangle = q_n|q_n\rangle$, sometimes had the same final incorrect answer as students who arbitrarily replaced $|\Psi\rangle$ with $|q_n\rangle$. For example, one student incorrectly reasoned that $\hat{Q}|\Psi\rangle = q_n|q_n\rangle$ and wrote $\langle\Psi|\hat{Q}|\Psi\rangle = \langle\Psi q_n|q_n\rangle = q_n\langle\Psi|q_n\rangle$. Another student who claimed that $|\Psi\rangle = |q_n\rangle$ wrote: "$\langle\Psi|\hat{Q}|\Psi\rangle = \langle\Psi|(\hat{Q}|q_n\rangle) = \langle\Psi|(q_n|q_n\rangle) = q_n\langle\Psi|q_n\rangle$ and $\langle\Psi|q_n\rangle$ is the component $q_n$ in $\Psi$." These two students had the same final answer despite their different reasoning. In interviews and some written responses, it was clear whether a student claimed that the operator $\hat{Q}$ acting on $|\Psi\rangle$ yields, e.g., $\hat{Q}|\Psi\rangle = q_n|q_n\rangle$, or arbitrarily replaced $|\Psi\rangle$ with $|q_n\rangle$. However, since the two difficulties can lead to the same final answer in written responses, it was sometimes unclear as to which category to code the difficulty. Thus, the two difficulties were combined into one category.

**Writing an incorrect expression for the expansion of $|\Psi\rangle$ using a complete set of eigenstates of the operator $\hat{Q}$.** Some students wrote incorrect expansions of $|\Psi\rangle$, e.g., $|\Psi\rangle = \sum_n |q_n\rangle$ or $|\Psi\rangle = \sum_n q_n |q_n\rangle$. Table III should be consulted for the specific percentages of students displaying this

difficulty. For example, one student stated: "$|\Psi\rangle$ can be expanded as a sum of eigenstates of $\hat{Q}$, $|\Psi\rangle = \sum_n q_n |q_n\rangle$." This student incorrectly claimed that the eigenvalues $q_n$ of the operator $\hat{Q}$ were the expansion coefficients $c_n$ when $|\Psi\rangle$ is expanded in terms of a complete set of eigenstates $|q_n\rangle$. If this were the case, the expansion coefficients will always be the same regardless of what $|\Psi\rangle$ actually is. Another student reasoned that $|\Psi\rangle = \sum |q_n\rangle$ and wrote: $\langle\Psi|\hat{Q}|\Psi\rangle = \sum\langle\Psi|\hat{Q}|q_n\rangle = \sum\langle\Psi|q_n|q_n\rangle = \sum q_n\langle\Psi|q_n\rangle$. Interestingly, this student did not change the generic state $\langle\Psi|$ in the "bra" state. We note that this type of reasoning may have led some students to incorrectly select statement 3 in question Q1 in Table I. These types of difficulties demonstrate that students have some correct knowledge, for example, they know that one can write $|\Psi\rangle$ as a superposition of the eigenstates of a generic operator $\hat{Q}$ and use this linear superposition to find the expectation value $\langle\Psi|\hat{Q}|\Psi\rangle$. However, interviews suggest that they often struggle to determine the appropriate expansion of $|\Psi\rangle$ or the coefficients of the expansion partly because they do not have a conceptual understanding of what the expansion coefficients mean.

**Writing an incorrect expression for the expectation value:** Some students wrote an incorrect expression for the expectation value, e.g., $\langle q_n|\hat{Q}|\Psi\rangle$ in which the "bra" and "ket" states are not the same. Table III shows the specific percentages of students displaying this difficulty in the open-ended questions. For example, one student wrote: $\langle q_n|\hat{Q}|\Psi\rangle = \sum_{n=1}^{\infty} q_n \Psi(q)$. These types of difficulties indicate that many students are not aware of the fact that the expectation value is found by "sandwiching" the operator between the "bra" and "ket" states in which the expectation value is evaluated, i.e., $\langle\Psi|\hat{Q}|\Psi\rangle$.

**Attempting to use the identity operator but getting lost along the way:** Some students were aware of the fact that one could find the expectation value by inserting the identity operator in the expression for the expectation value, but they had difficulty with the procedure and/or got lost along the way. Table III should be consulted for the specific percentages of students displaying these difficulties. For example, one common difficulty was an inability to distinguish between identity and projection operators. Other students had the correct expression for the identity operator but were unable to determine the expectation value correctly. For example, one student wrote the following: $\hat{I} = \sum_n |q_n\rangle\langle q_n|$, $\langle\Psi|\hat{Q}|\Psi\rangle = \sum\langle\Psi|\hat{Q}|q_n\rangle\langle q_n|\Psi\rangle = \sum\langle\Psi|\hat{Q}|q_n\rangle\Psi(q_n)$. This student was able to correctly insert the identity operator but did not define $\Psi(q_n)$ and left his final answer in terms of the operator $\hat{Q}$. Another student wrote: $\hat{I} = \sum_n|q_n\rangle\langle q_n|$, $\hat{I} = \int|q_n\rangle\langle q_n|dq$ and $\langle\Psi|\hat{Q}|\Psi\rangle = \langle\Psi|\hat{Q}\int|q\rangle\langle q|dq\,|\Psi\rangle = \langle\Psi|\int \hat{Q}|q\rangle\langle q|\Psi\rangle dq = \langle\Psi|\int q\langle q|\Psi\rangle dq$. This student was able to correctly insert the identity operator, but then stated that $\hat{Q}|q\rangle = q$ without the state $|q\rangle$ on the right hand side. Similar difficulties have been observed in other contexts as well. For example, prior research shows that some students believe that the Hamiltonian operator $\hat{H}$ acting on an energy eigenstate $|\psi_n\rangle$ yields the corresponding eigenvalue $E_n$, i.e., $\hat{H}|\psi_n\rangle = E_n$ [4]. Other students believe that the position operator $\hat{x}$ acting on a position eigenstate $|x'\rangle$ yields eigenvalue $x'$, i.e., $\hat{x}|x'\rangle = x'$. Some students physically justify their incorrect responses of this type by claiming that the Hamiltonian operator acting on its eigenstate corresponds to the measurement of energy and should yield energy on the right hand side of the equation or that the position operator acting on its eigenstate corresponds to the measurement of position and should yield position on the right hand side.

## IV. SUMMARY

Upper-level UG and G students had many types of common difficulties with the expectation value of an operator $\hat{Q}$ in terms of the eigenvalues and eigenstates of $\hat{Q}$ (e.g., when given that the states $\{|q_n\rangle, n = 1,2,3 \ldots \infty\}$ are the eigenstates of an operator $\hat{Q}$ corresponding to a physical observable with discrete eigenvalues $q_n$). Few students were able to reason conceptually about the expectation value and students often had many procedural difficulties in determining the expectation value of an observable in terms of the complete set of orthonormal eigenstates and eigenvalues of the corresponding hermitian operator.

Students must be given opportunities to build a good conceptual understanding of the expectation values of observables as well as develop the ability to calculate the expectation values. Instructors and researchers can use the student difficulties with the expectation values found in this study as a guide in developing curricula and pedagogies to help advanced students in QM courses develop a robust understanding of the expectation value.

### ACKNOWLEDGEMENTS

We thank the NSF for award PHY-1505460.

[1] C. Singh, Am. J. Phys. **69**, 885 (2001); **76**, 277 (2008); **76**, 400 (2008); arXiv:1602.05655; arXiv:1602.05664.
[2] M. Wittmann et al., Am. J. Phys. **70**, 218 (2002).; arXiv:1509.07740; arXiv:1603.02948; arXiv:1603.06025
[3] D. Zollman et al., Am. J. Phys. **70**, 252 (2002); arXiv:1510.01319; arXiv:1510.01300; arXiv:1602.06666
[4] C. Singh and E. Marshman, PRST PER **11**, 020117 (2015); arXiv:1509.04084; arXiv:1510.01296.
[5] G. Zhu and C. Singh, Am. J. Phys. **79**, 499 (2011); **80**, 252 (2012); PRST PER **8**, 010117 (2012); **8**, 010118 (2012); **9**, 010101 (2013); arXiv:1602.05619.
[6] Brown et al. PRST PER, **12**, 010121 (2016); arXiv:1602.05720; arXiv:1602.05461; arXiv:1602.06374
[7] G. Passante et al. PRST PER **11**, 020111 (2015).
[8] E. Marshman and C. Singh, PRST PER **11**, 020119 (2015); Eur. J. Phys. **37**, 024001 (2016).
[9] L. Bao and H. Sadaghiani, PERC Proc. 2006.